\documentclass[
    twocolumn,
	prd,
	amssymb,
	preprintnumbers,
    superscriptaddress,
	secnumarabic,
	nofootinbib]{revtex4-1}

\pdfoutput=1

\usepackage{tikz-feynman}
\tikzfeynmanset{compat=1.0.0}
\usepackage{graphicx}
\usepackage{enumitem}
\usepackage{latexsym}
\usepackage{amsfonts}
\usepackage{amssymb}
\usepackage{color}
\usepackage{amsmath}
\usepackage{slashed}
\usepackage{dcolumn}
\usepackage{verbatim}
\usepackage{float}
\usepackage{multirow}
\usepackage{xspace}
\usepackage[normalem]{ulem}
\usepackage{url}
\usepackage[
pdfauthor={Nirmal Raj}]{hyperref}

\newcommand{\gsim}{\gtrsim}
\newcommand{\lsim}{\lesssim}
\newcommand{\ra}{\rightarrow}

\def\Oc{\mathcal{O}}

\newcommand{\acro}[1]{\textsc{\MakeLowercase{#1}}} 
\newcommand{\osn}{\oldstylenums}
\renewcommand{\tilde}{\widetilde} 

\newcommand{\beq}{\begin{equation}}
\newcommand{\eeq}{\end{equation}}
\newcommand{\bea}{\begin{eqnarray}}
\newcommand{\eea}{\end{eqnarray}}
\newcommand{\nn}{\nonumber}

\def\mZ{m_{\rm Z}}
\def\mAd{m_{A_{\rm D}}}
\def\xB{x_{\rm B}}
\def\thetaW{\theta_{\rm W}}
\def\epsW{\epsilon_{\rm W}}
\def\gammaD{A_{\rm D}}

\allowdisplaybreaks

\begin{document}

\title{Breaking up the Proton:  An Affair with Dark Forces}

\author{Graham D.~Kribs}
\email{kribs@uoregon.edu}
\affiliation{Institute for Fundamental Science and Department of Physics, 
University of Oregon, Eugene, OR 97403, USA}
\author{David McKeen}
\email{mckeen@triumf.ca}
\author{Nirmal Raj} \email{nraj@triumf.ca}
\affiliation{TRIUMF, 4004 Wesbrook Mall, Vancouver, BC V6T 2A3, Canada}

\date{\today}

\begin{abstract}

Deep inelastic scattering of $e^{\pm}$ off protons is sensitive to 
contributions from ``dark photon'' exchange.  Using \acro{HERA} data
fit to \acro{HERA}'s parton distribution functions, we obtain the
model-independent bound $\epsilon \lsim 0.02$ on the kinetic mixing
between hypercharge and the dark photon for dark photon masses
$\lsim 10$~GeV\@.  This slightly improves on the bound obtained
from electroweak precision observables.  For higher masses the limit
weakens monotonically; $\epsilon \lsim 1$ for a dark photon mass
of $5$~TeV\@.  Utilizing \acro{PDF} sum rules,
we demonstrate that the effects of the dark photon cannot
be (trivially) absorbed into re-fit \acro{PDF}s, and in fact lead to
non-\acro{DGLAP} (Bjorken $\xB$-independent)
scaling violations that could provide a smoking gun in data. 
The proposed $e^\pm p$ collider operating
at $\sqrt{s} = 1.3$~TeV, \acro{lh}e\acro{c}, is anticipated to 
accumulate $10^3$ times the luminosity of \acro{HERA}, providing
substantial improvements in probing the effects of a dark photon:
sensitivity to $\epsilon$ well below that probed by electroweak
precision data is possible throughout virtually the entire dark
photon mass range, as well as being able to probe to much higher
dark photon masses, up to $100$~TeV\@.
\end{abstract}

\maketitle

{\bf \emph{Introduction.}}

Are there new gauge interactions in Nature?
A new, massive abelian vector boson (``dark photon'') can, at the renormalizable level, mix kinetically with the Standard Model hypercharge boson~\cite{Okun:1982xi,*Galison:1983pa,*Holdom:1985ag}:
\begin{equation}
 \mathcal{L} \supset
\frac{\epsilon}{2\cos\thetaW} F'_{\mu\nu}B^{\mu\nu}~.
\label{eq:Lag}
\end{equation}
Kinetic mixing with the hypercharge gauge boson becomes,
after electroweak symmetry breaking, mixing of the dark photon
with the neutral weak boson of the Standard Model (\acro{sm}).
We denote the unmixed dark photon by $A'_\mu$ and the unmixed
neutral weak boson by $\bar{Z}$. Diagonalizing the Lagrangian kinetic terms and gauge boson mass
matrix results in three physical vectors that couple to \acro{SM} fermions:
the massless photon $\gamma$, and the mass eigenstates $Z$ and $\gammaD$.

Numerous searches for the dark photon have been undertaken by directly producing it, in which case the signature depends on its decay mode. In the minimal setup where the only relevant couplings come from Eq.~\eqref{eq:Lag}, the dark photon decays back into charged standard model states, e.g. lepton pairs, offering striking signatures. However, the coupling in Eq.~\eqref{eq:Lag} may serve as our portal to a hidden sector that contains the particle species of the enigmatic dark matter~\cite{Pospelov:2007mp,essig2013dark}, and in this case the dark photon might decay invisibly or in a more complicated way depending on the structure of the hidden sector. It is therefore desirable to have ``decay-agnostic'' bounds that are independent of the details of a hidden sector.

\begin{figure}
    \centering
     \includegraphics[width=0.4\textwidth]{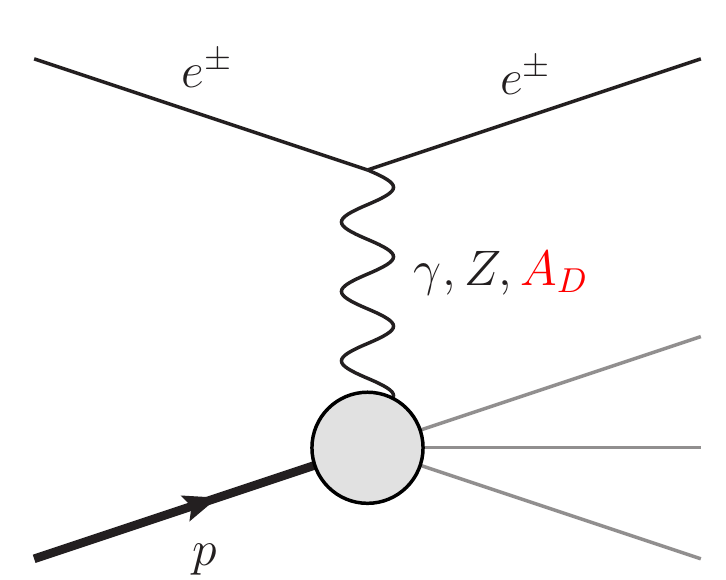} 
    \caption{Deep inelastic scattering of $e^\pm$ on the proton, mediated by the Standard Model photon and $Z$ boson, and a dark photon arising from kinetic mixing with an abelian hidden sector. 
    Measurements of this process at \acro{HERA} and \acro{LH}e\acro{C} probe the mixing parameter and dark photon mass without relying on any assumptions about the production and decay properties of $\gammaD$. }
    \label{fig:feyn}
\end{figure}

In this study, we investigate one such decay-agnostic process: deep inelastic scattering (\acro{dis}) of $e^{\pm}$ off protons. As seen in the Feynman diagram in Fig.~\ref{fig:feyn}, \acro{dis} in the presence of kinetic mixing is mediated by the photon, the $Z$, and the dark photon $\gammaD$. $\gammaD$ exchange leads to distinct non-\acro{DGLAP} scaling violations that may be constrained by existing data and may also be the smoking gun of a dark photon in future experiments.

Dark photon decay-agnostic limits on kinetic mixing were obtained in
Refs.~\cite{LEPModelIndep,Curtin:2014cca} from electroweak precision
observables (\acro{ewpo}), driven mainly by the $0.1\%$ precision
$Z$ pole-mass measurements at \acro{lep}.  
The main effect is a shift of the $Z$ mass relative to $m_{\rm W}/\cos\thetaW$, and using a global
fit to \acro{ewpo}, a bound of $\epsilon \lsim 0.03$ was obtained
for $\mAd \ll \mZ$.
We show that \acro{dis} measurements at the  $e^\pm p$ collider \acro{HERA}
can improve on this bound. 
With a net luminosity of 1 fb$^{-1}$, \acro{hera} achieved $1\%$
(systematics-limited) precision, however multiple measurements at this
precision give additional statistical power.
Decay-agnostic constraints also arise from measurements of the muon's
anomalous magnetic moment, which receives contributions from
$\gammaD$-mediated loop amplitudes~\cite{Pospelovgminus2};
these limits however weaken considerably for $\gammaD$ masses above
the muon mass, becoming negligible above $10$~GeV\@.
On the other hand, we show that \acro{DIS} can probe dark photon
masses of $10^4$~GeV and beyond.

As discussed above, if assumptions are made about the decay modes of the dark photon, 
additional constraints apply that may be considerably stronger
in the region $\mAd < 10$~GeV\@.
See Refs.~\cite{essig2013dark,serendip,fabbrichesi2020dark}
for a review of these constraints arising from colliders,
beam dump experiments, and other probes.\\

{\bf \emph{Signals of kinetic mixing in deep inelastic scattering.}}

\begin{figure}
    \centering
    \includegraphics[width=0.46\textwidth]{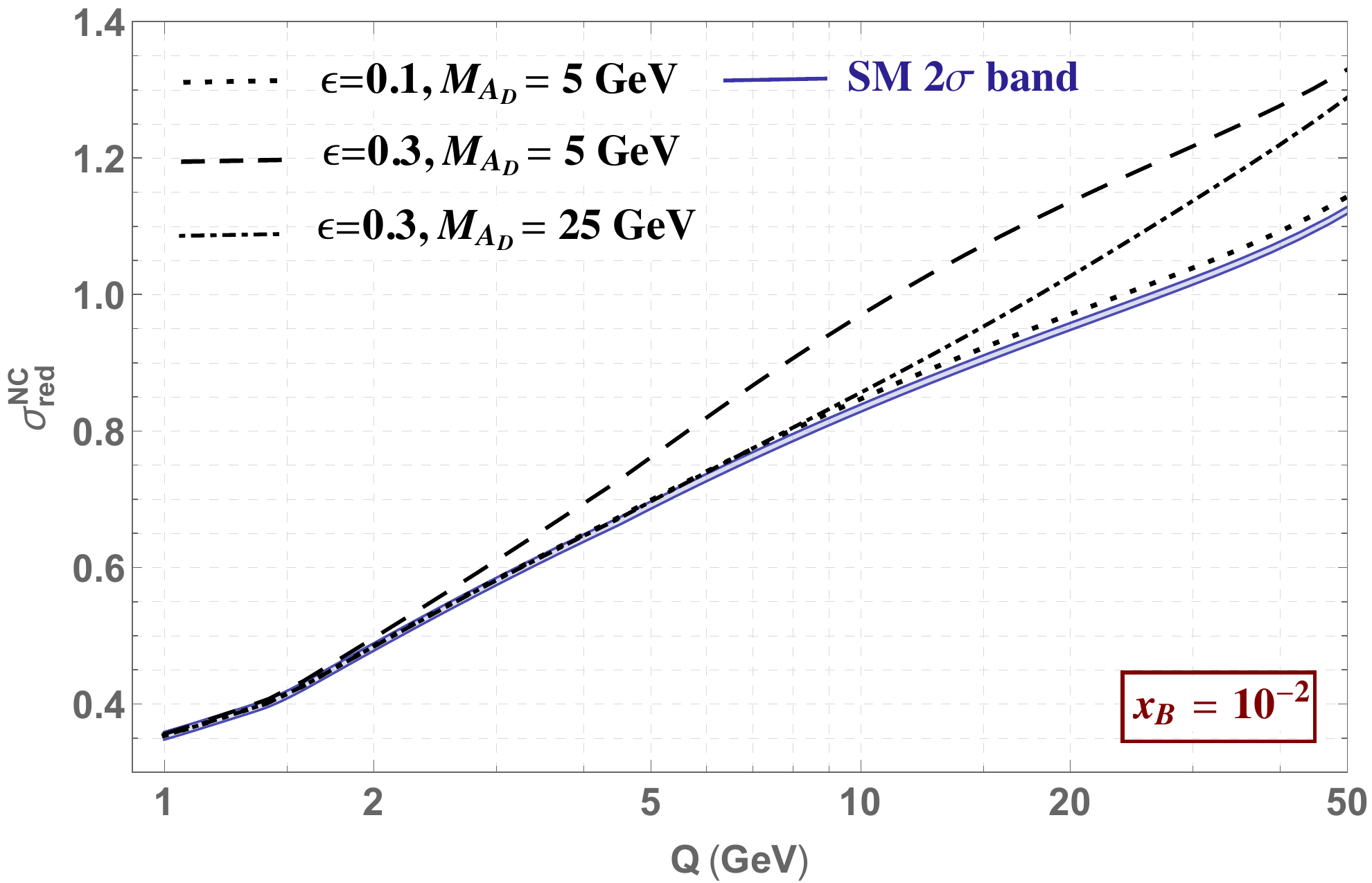} 
    \caption{
     \acro{DIS} neutral-current reduced cross sections for a representative Bjorken variable $\xB = 10^{-2}$, as estimated in Eq.~\eqref{eq:funcstruc} using \acro{HERAPDF}\osn{2}.\osn{0} \acro{LO}. 
     The blue band covers 2$\sigma$ uncertainties in the \acro{sm} cross section, obtained by using \acro{PDF} uncertainties and summing in quadrature the terms in Eq.~\eqref{eq:funcstruc}.
     In this regime where $Q \ll \mZ$, the effects of $Z$ are negligible and $\gammaD$ behaves like a massive photon constructively interfering with the \acro{sm} photon, leading to cross sections scaled by a factor of $1+2 \epsilon^2 Q^2/(Q^2+\mAd^2)$ relative to the \acro{SM}\@.  Importantly,
  the cross sections at all Bjorken $\xB$ values experience a shift \emph{at the same}
     $|Q| \sim \mAd$ that is a non-\acro{DGLAP} scaling violation of the cross sections, providing a prominent feature of the contribution from a dark photon.  
     Already  the $\epsilon = 0.1$ curve can be visually distinguished from the \acro{sm}, and when combined with the plethora of measurements at other $\xB$ values, it is clear that \acro{dis} at \acro{hera} can probe the considerably smaller values of $\epsilon \simeq 0.02$ as we show in Fig.~\ref{fig:ep}.
    }
    \label{fig:signals}
\end{figure}

In this section we review the basics of deep inelastic scattering,
incorporating dark photon exchange (for reviews, see e.g.\
\cite{review_dis,Abramowicz:2015mha, Kovarik:2019xvh}).
\acro{DIS} is described by the Lorentz-invariant kinematic variables
\begin{equation}
Q^2 = -q^2~, \ \xB = \frac{Q^2}{2q\cdot p}~, \ y = \frac{q\cdot p}{k\cdot p}~,
\end{equation}
where $q$ is the momentum transfer and $p$ ($k$) is the incoming proton's (electron's) momentum.
Using these kinematic variables, the unpolarized neutral-current
differential cross section rescaled as a (dimensionless)
``reduced cross section'' $\sigma^{\rm NC}_{\rm red}$ is:
\begin{equation}
\sigma^{\rm NC}_{\rm red} = \frac{Q^4\xB}{2\pi\alpha^2 [1 + (1-y)^2]}\frac{d^2\sigma}{d\xB dQ^2}~.
\end{equation}
The cross section can be expressed in terms of the parton distribution functions (\acro{pdf}s)
 as per the \acro{qcd} factorization theorem. 
In the Quark Parton Model (\acro{qpm}), \acro{dis} proceeds via elastic scattering on point-like quarks and anti-quarks, hence their \acro{pdf}s $f_q$ alone contribute and the variable $\xB$ becomes the momentum fraction of the proton carried by the parton in the infinite momentum frame; moreover longitudinal effects are negligible. 
Neglecting parity-violating effects, $\sigma^{\rm NC}_{\rm red}$ is equal to the structure function:
\begin{equation}  
\tilde{F}_2 = \sum_{i,j = \gamma,Z,\gammaD} \kappa_i \kappa_j F_2^{ij}~,
\label{eq:funcstruc}
\end{equation}
where
$\kappa_i = Q^2/(Q^2+M^2_{V_i})$ accounts for the propagators of vector bosons of mass $M_{V_i}$.  At leading order\footnote{By using the \acro{qcd} factorization theorem at next-to-leading order (\acro{nlo}) and by using \acro{nlo} \acro{PDF}s derived by \acro{hera}, we have checked that treating the problem at \acro{nlo} gives quite similar results. 
As discussed in Ref.~\cite{hera_combo}, while fitting \acro{pdf}s their uncertainties remain roughly the same as one goes to higher orders in $\alpha_s$.} in $\alpha_s$,
\begin{eqnarray}
\nn F_2^{ij} = \sum_{q} \xB f_q 
 (C^{\rm v}_{i,e}  C^{\rm v}_{j,e} + C^{\rm a}_{i,e}  C^{\rm a}_{j,e})
 (C^{\rm v}_{i,q}  C^{\rm v}_{j,q} + C^{\rm a}_{i,q}  C^{\rm a}_{j,q} )~,
\end{eqnarray}
where the summation runs over $q = u, \bar{u}, d, \bar{d}, c, \bar{c}, s, \bar{s}, b, \bar{b}$, and the vector and axial couplings to fermions (in units of $e=\sqrt{4\pi\alpha}$) are given as follows:
For the \acro{sm} photon, 
\begin{equation}
\{ C^{\rm v}_{\gamma,e},  C^{\rm v}_{\gamma,u},  C^{\rm v}_{\gamma,d}\} = \left \{ -1, \frac{2}{3}, -\frac{1}{3} \right \}, \ \   C^{\rm a}_\gamma = 0~.
\end{equation}
For the unmixed $\bar{Z}$ boson,
\begin{equation}
\bar{C}^{\rm v}_Z \sin 2\theta_{\rm W} = T_3^f - 2 q_f \sin^2 \theta_{\rm W}, \ \ \ \bar{C}^{\rm a}_Z \sin 2\theta_{\rm W} =  T_3^f~,
\end{equation}
where $\{ T_3^e, T_3^u, T_3^d \} = \{-1/2,1/2,-1/2\}$ is the weak isospin, $\{ q_e, q_u, q_d \} = \{ -1, 2/3, -1/3 \}$ is the electric charge, and the usual  Weinberg angle $\sin^2\thetaW \simeq 0.23127$~\cite{PDG}.

We can now add the effects of dark photon exchange. 
First we diagonalize the $A'$ mixing with hypercharge shown in Eq.~\eqref{eq:Lag} through the field redefinition: 
\begin{equation}
  B_\mu \rightarrow  B_\mu 
  + \frac{\epsilon}{\cos\theta_W} A'_\mu~,
\end{equation}
and canonically normalize the resulting dark photon kinetic term
through the field rescaling:
\begin{equation}
  A'_\mu \rightarrow  \frac{A'_\mu}{\sqrt{1-\epsilon^2/\cos^2\thetaW}}~.
\end{equation}
In the $\epsilon \ra \cos\thetaW$ limit the rescaling of $A'_\mu$ results in the enhancement of the dark gauge coupling, simultaneously enhancing couplings to \acro{sm} fermion currents. 
This in turn increases our sensitivity to large $\mAd$.

The dark photon and $\bar{Z}$ squared mass matrix becomes
\begin{equation}
M^2 =
\bar{m}_{\bar{Z}}^2
\begin{bmatrix}
1 & -\epsW \\
-\epsW & \epsW^2 + \rho^2
\end{bmatrix}~,
\label{eq:massmatrix}
\end{equation}
where 
\begin{equation}
\begin{aligned}
  \epsW &= \frac{\epsilon \tan\thetaW}{\sqrt{1-\epsilon^2/\cos^2\thetaW}}~,
  \\
  \rho &= \frac{\bar{m}_{A'}/\bar{m}_{\bar{\rm Z}}}{\sqrt{1-\epsilon^2/ \cos^2\thetaW}}~,
  \end{aligned}
  \label{eq:ratios}
\end{equation}
and the $\bar{Z}$-$A'$ mixing angle is given by
\begin{eqnarray}
\nn \tan \alpha &=& \frac{1}{2\epsW}
\bigg[ 1-\epsW^2-\rho^2 \\
\nn && -{\rm sign}(1-\rho^2)\sqrt{4\epsW^2+(1-\epsW^2-\rho^2)^2} \bigg]~. \\
\label{eq:mixang}
\end{eqnarray}

The physical $Z$ couplings are
\begin{eqnarray}
\nn C^{\rm v}_Z &=& (\cos \alpha - \epsW \sin \alpha) \bar{C}^{\rm v}_Z  + \epsW \sin \alpha \cot \thetaW C^{\rm v}_\gamma~,  \\
C^{\rm a}_Z &=& (\cos \alpha - \epsW \sin \alpha) \bar{C}^{\rm a}_Z~,
\label{eq:coupsZ}
\end{eqnarray}
while those of the physical $\gammaD$ are
\begin{eqnarray}
\nn C^{\rm v}_{A_{\rm D}} &=&  - (\sin \alpha + \epsW \cos \alpha) \bar{C}^{\rm v}_Z + \epsW \cos \alpha \cot \thetaW C^{\rm v}_\gamma~,  \\
C^{\rm a}_{A_{\rm D}} &=& - (\sin \alpha + \epsW \cos \alpha) \bar{C}^{\rm a}_Z~.
\label{eq:coupsAD}
\end{eqnarray}
The masses of the physical states are
\begin{equation}
\begin{aligned}
    m_{Z,\gammaD}^2&=\frac{m_{\bar Z}^2}{2}\bigg[1 + \epsW^2 + \rho^2
    \\
    &\pm{\rm sign}\left(1-\rho^2\right)\sqrt{\left(1 + \epsW^2 + \rho^2\right)^2-4\rho^2}\bigg].
\end{aligned}
\end{equation}
Note that for fixed $\epsilon$ and any value of $\bar{m}_{A'}/\bar{m}_{\bar{\rm Z}}$ the difference between the $Z$ and $\gammaD$ masses is always finite, $\left|m_{Z}^2-m_{\gammaD}^2\right|\geq 2\left|\epsW\right|m_{\bar Z}^2$. 
This ``eigenmass repulsion'', a well-known property of real symmetric matrices (including $M^2$) that implies that there are regions of the $\epsilon$-$m_{\gammaD}$ plane that cannot be realized. 

Note that the cross section in Eq.~\eqref{eq:funcstruc} is invariant under $\epsilon \ra -\epsilon$.
This arises from requiring $A'$ to couple to both quark and lepton currents to be observable at \acro{dis}, so that deviations from the \acro{sm} cross section arise first at $\mathcal{O}(\epsilon^2)$.

For $Q^2 \ll \mZ^2$, the short-distance $Z$-exchange is negligible, and $\gammaD$ modifies the \acro{DIS} cross section mainly through its constructive interference with the \acro{sm} photon.
Thus it effectively rescales the cross section by a factor of $[1+ \epsilon^2 Q^2/(Q^2+\mAd^2)]^2$ in this regime.
We illustrate this in Fig.~\ref{fig:signals} where we plot, for a representative $\xB = 10^{-2}$, $\sigma^{\rm NC}_{\rm red}$ versus $Q$ for $\epsilon = 0$ as a band covering $2\sigma$ uncertainties, and for 
($\epsilon$, $\mAd$/GeV) = 
(0.1, 5),  
(0.3, 5), and 
(0.3, 25).
Clearly larger $\epsilon$ values produce larger effects; more subtly, $\mAd$ sets the scale in $Q$ above which the effects of $\gammaD$ become significant.
Note that the $\epsilon = 0.1$ curve lies well outside the \acro{sm} band, indicating that \acro{hera} can probe $\epsilon \ll 0.1$ with a dataset spanning multiple $\xB$.

For $Q \gg \mZ$ \acro{DIS} probes the regime of unbroken electroweak symmetry, where the \acro{sm} process transpires effectively via massless $B$ exchange.
As we will see in the next section, here too the effect of kinetic mixing is to rescale the cross section by $[1+ \epsilon^2 Q^2/(Q^2+\mAd^2)]^2$. \\

{\bf \emph{HERA constraints and LHeC sensitivities.}}

To constrain kinetic mixing with \acro{dis}, we use 
the combined datasets of Runs I and II at \acro{hera}~\cite{hera_combo} over the ranges 
\begin{equation}
\nn 0.15 \leq Q^2/{\rm GeV}^2  \leq 3 \times 10^4~, \  5 \times 10^{-6} \leq \xB  \leq 0.65~.
\end{equation}
In principle, our constraints must be obtained fitting the \acro{hera} data simultaneously to both the dark photon parameters $(\epsilon,\mAd)$ and the \acro{PDF}s $f_q$.
In practice, however, we only fit to the dark photon parameters,\footnote{Our procedure can be viewed as estimating the expected sensitivities on our model using a second dataset with the same $(Q^2,\xB)$ grid as \acro{HERA}, with \acro{PDF}s fitted under the Standard Model null hypothesis. 
This is precisely what we do when we estimate the sensitivity of \acro{LH}e\acro{C} below.} and use the \acro{HERAPDF}\osn{2}.\osn{0} \acro{LO} \acro{pdf} set derived in Ref.~\cite{hera_combo} (importing it into Mathematica via ManeParse2.0~\cite{ManeParse}).

We believe the bounds we obtain from this simplified approach are,
in fact, robust against performing a simultaneous fit.  Let us justify this.  
First consider the region $\mAd \ll Q_{\rm min}$,
where $Q_{\rm min}$ is the smallest $Q$ probed at \acro{HERA}. 
The cross sections in Eq.~\eqref{eq:funcstruc} are rescaled by
$(1+\epsilon^2)^2$ with respect to the \acro{sm}, as discussed in the
previous section.
This implies that the \acro{PDF}s could absorb this rescaling of the cross
section by a simultaneous rescaling of all quark flavors, c.f. 
Eq.~\eqref{eq:funcstruc}.
However the normalization of $f_q$ is constrained by \acro{PDF} sum rules.
The quark-number sum rules
\begin{equation*}
 \int d\xB \left[f_q(\xB) - f_{\bar{q}}(\xB)\right] = 
 \begin{cases}
 2,~ q = u~, \\
 1,~q = d~, \\
 0,~q = s, c, b~ \, ,
 \end{cases}
\end{equation*}
and the momentum sum rule
\begin{equation}
\int d\xB~ \xB~ \left[\sum_q f_q (\xB) + f_g (\xB)\right] = 1 \, ,  
\end{equation}
applied over the \acro{HERA} $Q$ range are satisfied to
$\Oc(10^{-4})$ precision.
Using additional data in ranges of $Q^2$ outside \acro{hera}'s,
such as from beam dumps and hadron colliders, the sum rules
can be further constrained once \acro{DGLAP} evolution
is accounted for.  

\begin{figure}
    \centering
      \includegraphics[width=0.46\textwidth]{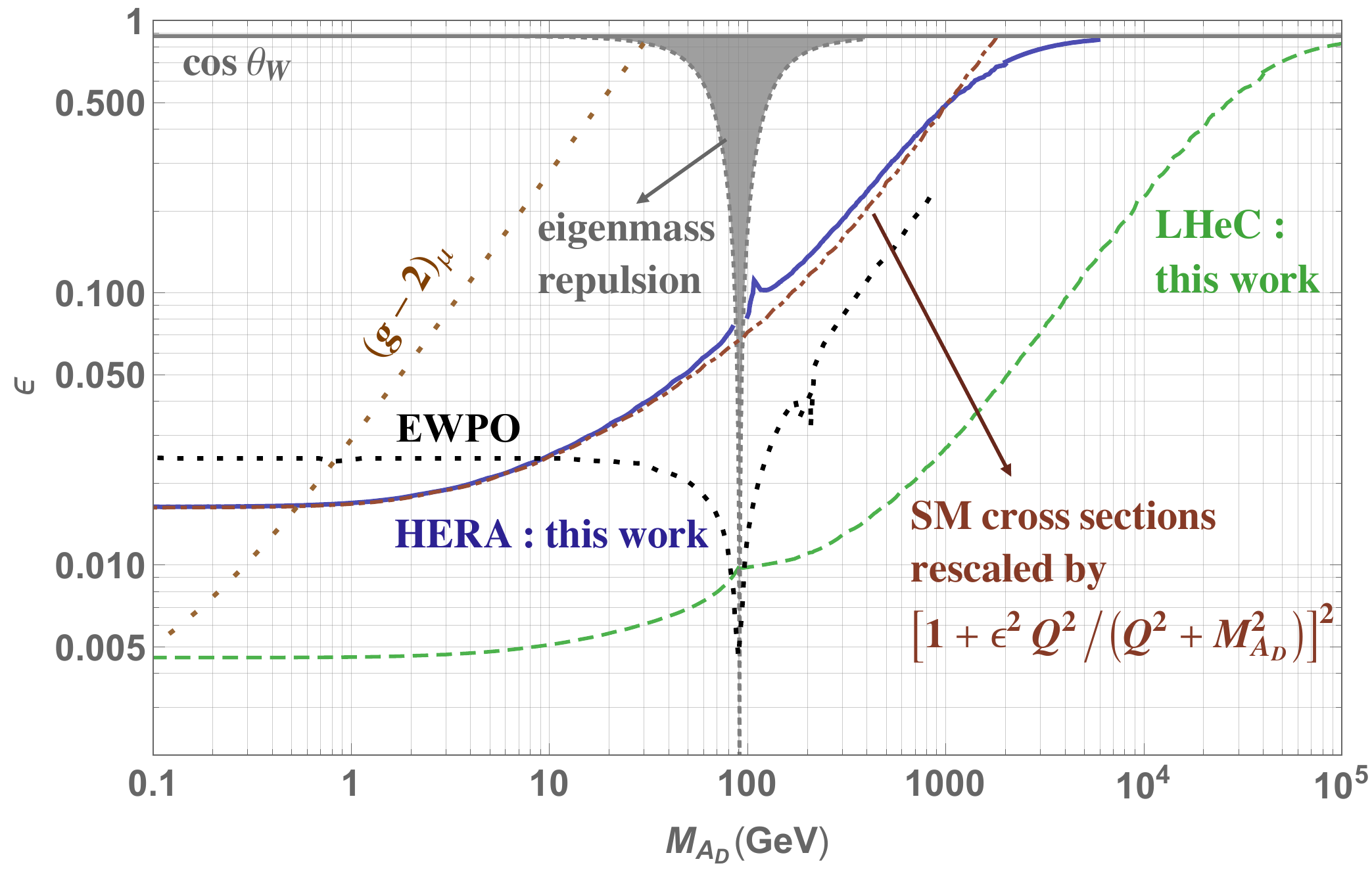} 
    \caption{
      95\% \acro{c.l.}\ limits on the kinetic mixing parameter vs dark photon mass from deep inelastic scattering (\acro{dis}) measurements. 
      Shown are limits from \acro{HERA} derived using \acro{HERAPDF}\osn{2}.\osn{0} \acro{LO} sets, and future sensitivities at \acro{lh}e\acro{c}, using \acro{PDF4LHC15}\_\acro{nnlo}\_\acro{lhec} \acro{pdf}s.
        For comparison are shown other decay-agnostic limits from measurements of electroweak precision observables (\acro{EWPO}) and the muon $g-2$.
        Also shown are hypothetical limits obtained by rescaling \acro{sm} \acro{dis} cross sections by a factor of $[1+\epsilon^2 Q^2/(Q^2+\mAd^2)]^2$, amounting to accounting only for interference between $\gammaD$ and $B$ boson exchange.
          In the gray-shaded region there is no physical value of $\mAd$ in the neighborhood of $\mZ$~=~91.1876~GeV due to repulsion of eigenmasses.
          The change in slope shift of the \acro{HERA} and \acro{LH}e\acro{C} sensitivity curves at large $\epsilon \gsim 0.7$ and $\mAd \gsim 1$~TeV occurs due to a factor of $1/\sqrt{1-\epsilon^2/\cos\thetaW^2}$ enhancement in the dark photon-fermion coupling. 
       See text for further details.
       }
    \label{fig:ep}
\end{figure}

Next, consider the parameter region where $\mAd$ is large compared
with the momentum exchange for \acro{DIS}.  
In this regime, the dark photon can be integrated out,
resulting in a rescaled cross section
$[1 +  \epsilon^2 Q^2/(Q^2 + \mAd^2)]^2 \simeq 1 + 2 \epsilon^2 Q^2/\mAd^2$.
The $Q^2$ polynomial growth in the rescaled cross section
leads to the largest corrections near $Q^2 \sim \mAd^2$, i.e.
near the edge of validity of the effective field theory. 
Reference~\cite{NPproton} showed that integrating out new particles
at $250$~GeV that led to modified quark-lepton interactions
could be mostly disentangled
from the logarithmic scaling of \acro{DGLAP} evolution.
That is, the new physics effects could not be easily
``fitted away'' into the \acro{PDF}s. 

Finally, for $Q_{\rm min} < \mAd < Q_{\rm max}$, there is a
(smoothed out) step in the rescaled cross section at $Q\simeq \mAd$ which comes from the factor of
$[1 +  \epsilon^2 Q^2/(Q^2 + \mAd^2)]^2$, illustrated in Fig.~\ref{fig:signals}.
Since this step occurs at the same $Q^2$ for all $\xB$ values,
it also does not behave like \acro{DGLAP} evolution.
 
In order to obtain the net uncertainty in $\sigma^{\rm NC}_{\rm red}$,
we use the \acro{herapdf} uncertainties to sum in quadrature the uncertainties of the terms in Eq.~\eqref{eq:funcstruc}.
It is these uncertainties that we will use to estimate our limits, as opposed to the errors in the ``raw'' measurements of the cross sections, since (a) the covariance matrix describing correlations among the errors is not given, and (b) the \acro{pdf} fitting procedure accounts for these correlations.
 
In deriving our bounds, we use the ($Q^2,\xB$) grid used for the \acro{hera} run involving $e^+$ scattering with $\sqrt{s} = 318$~GeV and $0.5$~fb$^{-1}$ luminosity, as given in Table~10 of Ref.~\cite{hera_combo}.
 This grid, containing 485 points, covers most of the ($Q^2,\xB$) used in the other runs involving $e^\pm$ scattering at smaller $\sqrt{s}$ and luminosity; although data from all these runs were used for fitting \acro{pdf}s, we do not use these other grids to avoid oversampling.
 We derive the \osn{95}\% \acro{c.l.} limit by locating values of ($\epsilon,\mAd$) for which
 \begin{equation}
 \chi^2 = \sum_{\rm grid} \frac{(\sigma^{\rm NC}_{\rm red} - \sigma^{\rm NC}_{\rm red}|_{\epsilon \ra 0})^2}{(\delta \sigma^{\rm NC}_{\rm red})^2} = 5.99~,
 \end{equation}
where the summation is over the ($Q^2,\xB$) grid mentioned above.
The resulting limits are displayed in Fig.~\ref{fig:ep}.  We also show the decay-agnostic limits from the $(g-2)_\mu$ measurement at the E821 experiment, requiring 5$\sigma$ deviation from the central value using the calculations in Ref.~\cite{Pospelovgminus2}, as well as the limits from \acro{ewpo} derived in Ref.~\cite{LEPModelIndep}.
Our bounds are driven largely by about 25 data points in the  ($Q^2,\xB$) grid where the cross section is obtained with a maximum precision of 0.3\%-0.4\%.
This is the origin of why our limits are (slightly) stronger than \acro{ewpo} for $\mAd \lsim 10$~GeV~$\ll \mZ$. 
In this regime the observable correction at both \acro{lep} $Z$ pole measurements and \acro{dis} scales as $\epsilon^2$, and while \acro{lep} operated at a precision of 0.1\%, which appears better than our precision, our bound actually benefits from 25 independent measurements, effectively diminishing our uncertainty by a statistical factor of $\sqrt{25}$.
We also display a hypothetical bound obtained by simply rescaling the \acro{sm} cross sections by $1+2 \epsilon^2 Q^2/(Q^2+\mAd^2)$, seen to trail the actual bound with amusing proximity.
As discussed earlier, such a rescaling amounts to accounting only for $\gammaD$ interference with the $B$ exchange amplitude: for $Q \ll \mZ$ this effectively rescales photon exchange, and for higher $Q$ it rescales both photon and $Z$ exchange.

In this figure we also show the \osn{95}\% \acro{c.l.}\ sensitivity of the future $e^\pm p$ collider \acro{LH}e\acro{C}~\cite{LHeCDesign} derived by using the ($Q^2,\xB$) grid for $e^+$ scattering over the range  
\begin{equation}
\nn 5 \leq Q^2/{\rm GeV}^2 \leq 10^6~, \ 5 \times 10^{-6} \leq \xB \leq 0.8~ \, .
\end{equation}
The \acro{lh}e\acro{c} is anticipated to obtain $10^3$ times the integrated luminosity of \acro{hera}, thus gaining in statistical precision by a factor of about $30$. 
 We are interested in characterizing the maximal sensitivity that \acro{lh}e\acro{c} could achieve with this increased precision.  This is a different objective from obtaining the best-fit \acro{PDF}s across all datasets. 
 Therefore, to estimate \acro{lh}e\acro{c} sensitivity, we use
\acro{PDF4LHC15}\_\acro{nnlo}\_\acro{lhec} \acro{pdf}s fitted to pseudo-data  Ref.~\cite{LHeCPDFs}, but then rescale the fractional uncertainties to match with $\sqrt{1/10^3}$ times the fractional uncertainties of \acro{HERAPDF}\osn{2}.\osn{0} \acro{LO}, optimistically assuming that systematic errors can be kept below this level.  We have checked that 
rescaling the $Q^2$ values of the \acro{hera} grid by a factor of $5/0.15 = 10^6/(3 \times 10^4)$, the envelopes 
of smallest uncertainties (as a function of $Q^2$)
for either \acro{PDF} set are well-aligned.
We see from the figure that \acro{lh}e\acro{c} exceeds \acro{hera} in the entire $\mAd$ range constrained by the latter, and indeed reaches $\mAd$ up to $100$~TeV thanks to probing the proton at very high $Q$.

The \acro{HERAPDF}\osn{2}.\osn{0} \acro{LO} \acro{pdf} set is designed to fit solely the \acro{HERA} \acro{DIS} data.  We used this set not because we believe this is the best description of the quark \acro{pdf}s, but because this is the most accurate interpolated description of the \acro{HERA} data.  Since we are interested in the sensitivity of \acro{HERA} alone to kinetic mixing, we believe this is the correct approach to obtain the most accurate estimate of the sensitivity.   We point out, however, that a more wide-ranging description of \acro{pdf}s require 
``global fits'' to \acro{dis} at \acro{hera} combined with beam-dump and hadron collider experiment datasets that include complementary ranges of $Q^2$ and $\xB$.
Such \acro{PDF} determinations contain additional sources of uncertainty \cite{Kovarik:2019xvh}:
(1) a ``tolerance'' factor to rescale the goodness-of-fit so that tensions in fitting multiple datasets may be eased to within 1$\sigma$ uncertainty,
(2) parameterization uncertainties introduced by the need to use numerous parameters to fit numerous datasets.
The combination of these effects significantly increases the \acro{pdf} uncertainties.
Indeed we find that, had we used the global \acro{pdf} set \acro{ct}\osn{18}\acro{z}~\cite{ct18}, our bounds on $\epsilon$ would be weakened by a factor of up to $3$. 
We do not believe this is a fair characterization of \acro{HERA}'s bounds on kinetic mixing.

Finally, we note that \acro{EWPO} sensitivities on $\epsilon$ are expected to improve by $\mathcal{O}(1)$ factors (a factor of $\sim$10) with increased sensitivities provided by future \acro{lhc} (\acro{ilc} in GigaZ mode) measurements~\cite{Curtin:2014cca}. \\

{\bf \emph{DIS-cussion.}}

Deep inelastic scattering of $e^\pm$ off protons is a sensitive,
model-independent probe of kinetic mixing with a dark photon up
to $100$~TeV masses.  No assumptions need to be made regarding
the dark photon's decay modes.
We find \acro{hera} data is slightly more sensitive than \acro{EWPO}
for dark photon masses less than about $10$~GeV\@.  The
\acro{LH}e\acro{C} could significantly improve the sensitivity
of \acro{dis} to kinetic mixing, probing values of $\epsilon$
well below the sensitivity of \acro{ewpo} data.  

It is intriguing to consider the possibility of \emph{discovering}
a dark photon's signature in \acro{DIS} data.  This seems quite unlikely
with existing \acro{HERA} data, since \acro{EWPO} leads to a stronger
constraint for most of the parameter space. The main constraint
from \acro{EWPO} arises due to a shift of $\mZ$ relative to
$m_{\rm W}/\cos\thetaW$.  It is possible, though unlikely, that other
physics in the dark sector could compensate for this apparent
contribution to custodial violation and weaken the \acro{EWPO} bounds.
In addition, the parameter region
$\mAd \lsim 10$~GeV where \acro{DIS} is slightly more sensitive
is strongly constrained by \emph{model-dependent}
searches, especially from $B$-factories.  
In particular, the BaBar collaboration has searched for dark photons produced via $e^+e^-\to\gamma \gammaD$ assuming that $\gammaD$ decays visibly through its kinetic mixing~\cite{Lees:2014xha} or invisibly into a dark sector~\cite{Lees:2017lec}. 
In both cases limits on $\epsilon$ at the level of $10^{-3}$ are obtained. 
A similar search by \acro{LHC}b in the $\mu^+\mu^-$ final state constrains dark photons to $\epsilon \lsim \Oc(10^{-3})$ up to 70~GeV masses~\cite{lhcb}.
These limits can potentially be weakened if $\gammaD$ couples to a dark sector with further structure, leading to more complicated final states as in, e.g.,~\cite{Baumgart:2009tn,*Essig:2009nc}.

The \acro{LH}e\acro{C}'s sensitivity is significantly better
than \acro{EWPO}, and this provides the most exciting possibility
to directly search for the non-\acro{DGLAP} (Bjorken $\xB$-independent) scaling
violation in the cross section 
illustrated in Fig.~\ref{fig:signals}.  Maximizing the sensitivity would
be best optimized by simultaneously fitting the \acro{PDF}s
with dark photon exchange.  Nevertheless, we have emphasized that
\acro{pdf} sum rules provide strong constraints on ``fitting away'' the
effects of a dark photon on \acro{PDF}s, and other studies~\cite{NPproton} have
also found that \acro{PDF}s do not easily fit away the polynomial
scaling exhibited by massive dark photons.  

In this study we have focused on the effects of a dark photon on
\acro{DIS}, however this can also be extended to any new force between
quarks and leptons, such as mediated by a gauged $U(1)_{B-L}$ vector
boson, new scalars in the Higgs sector, or other exotic force carriers.
We leave these investigations to future work. \\

{\bf \emph{Acknowledgments.}}
We are grateful to
Paddy Fox,
David Morrissey,
John Ng,
Maxim Pospelov,
Tim Tait, 
Yue Zhao,
and especially
Dave Soper 
for beneficial conversations.
The work of \acro{g.d.k.}\ is supported in part by the
U.S. Department of Energy under Grant Number DE-SC0011640.
The work of \acro{d.m.}\ and \acro{n.r.}\ is supported by the Natural Sciences and Engineering Research Council of Canada. 
T\acro{RIUMF} receives federal funding via a contribution agreement with the National Research Council Canada.
This work was performed in part at the Aspen Center for Physics, which is supported by National
Science Foundation grant PHY-1607611.

\bibliography{refs}

\begin{thebibliography}{25}%
\makeatletter
\providecommand \@ifxundefined [1]{%
 \@ifx{#1\undefined}
}%
\providecommand \@ifnum [1]{%
 \ifnum #1\expandafter \@firstoftwo
 \else \expandafter \@secondoftwo
 \fi
}%
\providecommand \@ifx [1]{%
 \ifx #1\expandafter \@firstoftwo
 \else \expandafter \@secondoftwo
 \fi
}%
\providecommand \natexlab [1]{#1}%
\providecommand \enquote  [1]{``#1''}%
\providecommand \bibnamefont  [1]{#1}%
\providecommand \bibfnamefont [1]{#1}%
\providecommand \citenamefont [1]{#1}%
\providecommand \href@noop [0]{\@secondoftwo}%
\providecommand \href [0]{\begingroup \@sanitize@url \@href}%
\providecommand \@href[1]{\@@startlink{#1}\@@href}%
\providecommand \@@href[1]{\endgroup#1\@@endlink}%
\providecommand \@sanitize@url [0]{\catcode `\\12\catcode `\$12\catcode
  `\&12\catcode `\#12\catcode `\^12\catcode `\_12\catcode `\%12\relax}%
\providecommand \@@startlink[1]{}%
\providecommand \@@endlink[0]{}%
\providecommand \url  [0]{\begingroup\@sanitize@url \@url }%
\providecommand \@url [1]{\endgroup\@href {#1}{\urlprefix }}%
\providecommand \urlprefix  [0]{URL }%
\providecommand \Eprint [0]{\href }%
\providecommand \doibase [0]{http://dx.doi.org/}%
\providecommand \selectlanguage [0]{\@gobble}%
\providecommand \bibinfo  [0]{\@secondoftwo}%
\providecommand \bibfield  [0]{\@secondoftwo}%
\providecommand \translation [1]{[#1]}%
\providecommand \BibitemOpen [0]{}%
\providecommand \bibitemStop [0]{}%
\providecommand \bibitemNoStop [0]{.\EOS\space}%
\providecommand \EOS [0]{\spacefactor3000\relax}%
\providecommand \BibitemShut  [1]{\csname bibitem#1\endcsname}%
\let\auto@bib@innerbib\@empty
\bibitem [{\citenamefont {Okun}(1982)}]{Okun:1982xi}%
  \BibitemOpen
  \bibfield  {author} {\bibinfo {author} {\bibfnamefont {L.}~\bibnamefont
  {Okun}},\ }\href@noop {} {\bibfield  {journal} {\bibinfo  {journal} {Sov.
  Phys. JETP}\ }\textbf {\bibinfo {volume} {56}},\ \bibinfo {pages} {502}
  (\bibinfo {year} {1982})}\BibitemShut {NoStop}%
\bibitem [{\citenamefont {Galison}\ and\ \citenamefont
  {Manohar}(1984)}]{Galison:1983pa}%
  \BibitemOpen
  \bibfield  {author} {\bibinfo {author} {\bibfnamefont {P.}~\bibnamefont
  {Galison}}\ and\ \bibinfo {author} {\bibfnamefont {A.}~\bibnamefont
  {Manohar}},\ }\href {\doibase 10.1016/0370-2693(84)91161-4} {\bibfield
  {journal} {\bibinfo  {journal} {Phys. Lett. B}\ }\textbf {\bibinfo {volume}
  {136}},\ \bibinfo {pages} {279} (\bibinfo {year} {1984})}\BibitemShut
  {NoStop}%
\bibitem [{\citenamefont {Holdom}(1986)}]{Holdom:1985ag}%
  \BibitemOpen
  \bibfield  {author} {\bibinfo {author} {\bibfnamefont {B.}~\bibnamefont
  {Holdom}},\ }\href {\doibase 10.1016/0370-2693(86)91377-8} {\bibfield
  {journal} {\bibinfo  {journal} {Phys. Lett.}\ }\textbf {\bibinfo {volume}
  {166B}},\ \bibinfo {pages} {196} (\bibinfo {year} {1986})}\BibitemShut
  {NoStop}%
\bibitem [{\citenamefont {Pospelov}\ \emph {et~al.}(2008)\citenamefont
  {Pospelov}, \citenamefont {Ritz},\ and\ \citenamefont
  {Voloshin}}]{Pospelov:2007mp}%
  \BibitemOpen
  \bibfield  {author} {\bibinfo {author} {\bibfnamefont {M.}~\bibnamefont
  {Pospelov}}, \bibinfo {author} {\bibfnamefont {A.}~\bibnamefont {Ritz}}, \
  and\ \bibinfo {author} {\bibfnamefont {M.~B.}\ \bibnamefont {Voloshin}},\
  }\href {\doibase 10.1016/j.physletb.2008.02.052} {\bibfield  {journal}
  {\bibinfo  {journal} {Phys. Lett.}\ }\textbf {\bibinfo {volume} {B662}},\
  \bibinfo {pages} {53} (\bibinfo {year} {2008})},\ \Eprint
  {http://arxiv.org/abs/0711.4866} {arXiv:0711.4866 [hep-ph]} \BibitemShut
  {NoStop}%
\bibitem [{\citenamefont {Essig~{\em et al}}(2013)}]{essig2013dark}%
  \BibitemOpen
  \bibfield  {author} {\bibinfo {author} {\bibfnamefont {R.}~\bibnamefont
  {Essig~{\em et al}}},\ }\href@noop {} {\enquote {\bibinfo {title} {Dark
  sectors and new, light, weakly-coupled particles},}\ } (\bibinfo {year}
  {2013}),\ \Eprint {http://arxiv.org/abs/1311.0029} {arXiv:1311.0029 [hep-ph]}
  \BibitemShut {NoStop}%
\bibitem [{\citenamefont {Hook}\ \emph {et~al.}(2011)\citenamefont {Hook},
  \citenamefont {Izaguirre},\ and\ \citenamefont {Wacker}}]{LEPModelIndep}%
  \BibitemOpen
  \bibfield  {author} {\bibinfo {author} {\bibfnamefont {A.}~\bibnamefont
  {Hook}}, \bibinfo {author} {\bibfnamefont {E.}~\bibnamefont {Izaguirre}}, \
  and\ \bibinfo {author} {\bibfnamefont {J.~G.}\ \bibnamefont {Wacker}},\
  }\href {\doibase 10.1155/2011/859762} {\bibfield  {journal} {\bibinfo
  {journal} {Adv. High Energy Phys.}\ }\textbf {\bibinfo {volume} {2011}},\
  \bibinfo {pages} {859762} (\bibinfo {year} {2011})},\ \Eprint
  {http://arxiv.org/abs/1006.0973} {arXiv:1006.0973 [hep-ph]} \BibitemShut
  {NoStop}%
\bibitem [{\citenamefont {Curtin}\ \emph {et~al.}(2015)\citenamefont {Curtin},
  \citenamefont {Essig}, \citenamefont {Gori},\ and\ \citenamefont
  {Shelton}}]{Curtin:2014cca}%
  \BibitemOpen
  \bibfield  {author} {\bibinfo {author} {\bibfnamefont {D.}~\bibnamefont
  {Curtin}}, \bibinfo {author} {\bibfnamefont {R.}~\bibnamefont {Essig}},
  \bibinfo {author} {\bibfnamefont {S.}~\bibnamefont {Gori}}, \ and\ \bibinfo
  {author} {\bibfnamefont {J.}~\bibnamefont {Shelton}},\ }\href {\doibase
  10.1007/JHEP02(2015)157} {\bibfield  {journal} {\bibinfo  {journal} {JHEP}\
  }\textbf {\bibinfo {volume} {02}},\ \bibinfo {pages} {157} (\bibinfo {year}
  {2015})},\ \Eprint {http://arxiv.org/abs/1412.0018} {arXiv:1412.0018
  [hep-ph]} \BibitemShut {NoStop}%
\bibitem [{\citenamefont {Pospelov}(2009)}]{Pospelovgminus2}%
  \BibitemOpen
  \bibfield  {author} {\bibinfo {author} {\bibfnamefont {M.}~\bibnamefont
  {Pospelov}},\ }\href {\doibase 10.1103/PhysRevD.80.095002} {\bibfield
  {journal} {\bibinfo  {journal} {Phys. Rev.}\ }\textbf {\bibinfo {volume}
  {D80}},\ \bibinfo {pages} {095002} (\bibinfo {year} {2009})},\ \Eprint
  {http://arxiv.org/abs/0811.1030} {arXiv:0811.1030 [hep-ph]} \BibitemShut
  {NoStop}%
\bibitem [{\citenamefont {Ilten}\ \emph {et~al.}(2018)\citenamefont {Ilten},
  \citenamefont {Soreq}, \citenamefont {Williams},\ and\ \citenamefont
  {Xue}}]{serendip}%
  \BibitemOpen
  \bibfield  {author} {\bibinfo {author} {\bibfnamefont {P.}~\bibnamefont
  {Ilten}}, \bibinfo {author} {\bibfnamefont {Y.}~\bibnamefont {Soreq}},
  \bibinfo {author} {\bibfnamefont {M.}~\bibnamefont {Williams}}, \ and\
  \bibinfo {author} {\bibfnamefont {W.}~\bibnamefont {Xue}},\ }\href {\doibase
  10.1007/JHEP06(2018)004} {\bibfield  {journal} {\bibinfo  {journal} {JHEP}\
  }\textbf {\bibinfo {volume} {06}},\ \bibinfo {pages} {004} (\bibinfo {year}
  {2018})},\ \Eprint {http://arxiv.org/abs/1801.04847} {arXiv:1801.04847
  [hep-ph]} \BibitemShut {NoStop}%
\bibitem [{\citenamefont {Fabbrichesi}\ \emph {et~al.}(2020)\citenamefont
  {Fabbrichesi}, \citenamefont {Gabrielli},\ and\ \citenamefont
  {Lanfranchi}}]{fabbrichesi2020dark}%
  \BibitemOpen
  \bibfield  {author} {\bibinfo {author} {\bibfnamefont {M.}~\bibnamefont
  {Fabbrichesi}}, \bibinfo {author} {\bibfnamefont {E.}~\bibnamefont
  {Gabrielli}}, \ and\ \bibinfo {author} {\bibfnamefont {G.}~\bibnamefont
  {Lanfranchi}},\ }\href@noop {} {\enquote {\bibinfo {title} {The dark
  photon},}\ } (\bibinfo {year} {2020}),\ \Eprint
  {http://arxiv.org/abs/2005.01515} {arXiv:2005.01515 [hep-ph]} \BibitemShut
  {NoStop}%
\bibitem [{\citenamefont {Blumlein}(2013)}]{review_dis}%
  \BibitemOpen
  \bibfield  {author} {\bibinfo {author} {\bibfnamefont {J.}~\bibnamefont
  {Blumlein}},\ }\href {\doibase 10.1016/j.ppnp.2012.09.006} {\bibfield
  {journal} {\bibinfo  {journal} {Prog. Part. Nucl. Phys.}\ }\textbf {\bibinfo
  {volume} {69}},\ \bibinfo {pages} {28} (\bibinfo {year} {2013})},\ \Eprint
  {http://arxiv.org/abs/1208.6087} {arXiv:1208.6087 [hep-ph]} \BibitemShut
  {NoStop}%
\bibitem [{\citenamefont {Abramowicz}\ \emph {et~al.}(2015)\citenamefont
  {Abramowicz} \emph {et~al.}}]{Abramowicz:2015mha}%
  \BibitemOpen
  \bibfield  {author} {\bibinfo {author} {\bibfnamefont {H.}~\bibnamefont
  {Abramowicz}} \emph {et~al.} (\bibinfo {collaboration} {H1, ZEUS}),\ }\href
  {\doibase 10.1140/epjc/s10052-015-3710-4} {\bibfield  {journal} {\bibinfo
  {journal} {Eur. Phys. J.}\ }\textbf {\bibinfo {volume} {C75}},\ \bibinfo
  {pages} {580} (\bibinfo {year} {2015})},\ \Eprint
  {http://arxiv.org/abs/1506.06042} {arXiv:1506.06042 [hep-ex]} \BibitemShut
  {NoStop}%
\bibitem [{\citenamefont {Kovařík}\ \emph {et~al.}(2019)\citenamefont
  {Kovařík}, \citenamefont {Nadolsky},\ and\ \citenamefont
  {Soper}}]{Kovarik:2019xvh}%
  \BibitemOpen
  \bibfield  {author} {\bibinfo {author} {\bibfnamefont {K.}~\bibnamefont
  {Kovařík}}, \bibinfo {author} {\bibfnamefont {P.~M.}\ \bibnamefont
  {Nadolsky}}, \ and\ \bibinfo {author} {\bibfnamefont {D.~E.}\ \bibnamefont
  {Soper}},\ }\href@noop {} {\  (\bibinfo {year} {2019})},\ \Eprint
  {http://arxiv.org/abs/1905.06957} {arXiv:1905.06957 [hep-ph]} \BibitemShut
  {NoStop}%
\bibitem [{\citenamefont {H1}\ and\ \citenamefont
  {Collaborations}(2015)}]{hera_combo}%
  \BibitemOpen
  \bibfield  {author} {\bibinfo {author} {\bibnamefont {H1}}\ and\ \bibinfo
  {author} {\bibfnamefont {Z.}~\bibnamefont {Collaborations}},\ }\href@noop {}
  {\enquote {\bibinfo {title} {Combination of measurements of inclusive deep
  inelastic $e^{\pm}p$ scattering cross sections and qcd analysis of hera
  data},}\ } (\bibinfo {year} {2015}),\ \Eprint
  {http://arxiv.org/abs/1506.06042} {arXiv:1506.06042 [hep-ex]} \BibitemShut
  {NoStop}%
\bibitem [{\citenamefont {Tanabashi}\ \emph {et~al.}(2018)\citenamefont
  {Tanabashi} \emph {et~al.}}]{PDG}%
  \BibitemOpen
  \bibfield  {author} {\bibinfo {author} {\bibfnamefont {M.}~\bibnamefont
  {Tanabashi}} \emph {et~al.} (\bibinfo {collaboration} {Particle Data
  Group}),\ }\href {\doibase 10.1103/PhysRevD.98.030001} {\bibfield  {journal}
  {\bibinfo  {journal} {Phys. Rev.}\ }\textbf {\bibinfo {volume} {D98}},\
  \bibinfo {pages} {030001} (\bibinfo {year} {2018})}\BibitemShut {NoStop}%
\bibitem [{\citenamefont {Clark}\ \emph {et~al.}(2017)\citenamefont {Clark},
  \citenamefont {Godat},\ and\ \citenamefont {Olness}}]{ManeParse}%
  \BibitemOpen
  \bibfield  {author} {\bibinfo {author} {\bibfnamefont {D.}~\bibnamefont
  {Clark}}, \bibinfo {author} {\bibfnamefont {E.}~\bibnamefont {Godat}}, \ and\
  \bibinfo {author} {\bibfnamefont {F.}~\bibnamefont {Olness}},\ }\href
  {\doibase 10.1016/j.cpc.2017.03.004} {\bibfield  {journal} {\bibinfo
  {journal} {Computer Physics Communications}\ }\textbf {\bibinfo {volume}
  {216}},\ \bibinfo {pages} {126–137} (\bibinfo {year} {2017})}\BibitemShut
  {NoStop}%
\bibitem [{\citenamefont {Carrazza}\ \emph {et~al.}(2019)\citenamefont
  {Carrazza}, \citenamefont {Degrande}, \citenamefont {Iranipour},
  \citenamefont {Rojo},\ and\ \citenamefont {Ubiali}}]{NPproton}%
  \BibitemOpen
  \bibfield  {author} {\bibinfo {author} {\bibfnamefont {S.}~\bibnamefont
  {Carrazza}}, \bibinfo {author} {\bibfnamefont {C.}~\bibnamefont {Degrande}},
  \bibinfo {author} {\bibfnamefont {S.}~\bibnamefont {Iranipour}}, \bibinfo
  {author} {\bibfnamefont {J.}~\bibnamefont {Rojo}}, \ and\ \bibinfo {author}
  {\bibfnamefont {M.}~\bibnamefont {Ubiali}},\ }\href {\doibase
  10.1103/physrevlett.123.132001} {\bibfield  {journal} {\bibinfo  {journal}
  {Physical Review Letters}\ }\textbf {\bibinfo {volume} {123}} (\bibinfo
  {year} {2019}),\ 10.1103/physrevlett.123.132001}\BibitemShut {NoStop}%
\bibitem [{\citenamefont {Abelleira~Fernandez}\ \emph
  {et~al.}(2012)\citenamefont {Abelleira~Fernandez}, \citenamefont {Adolphsen},
  \citenamefont {Akay}, \citenamefont {Aksakal}, \citenamefont {Albacete},
  \citenamefont {Alekhin}, \citenamefont {Allport}, \citenamefont {Andreev},
  \citenamefont {Appleby}, \citenamefont {Arikan},\ and\ \citenamefont
  {et~al.}}]{LHeCDesign}%
  \BibitemOpen
  \bibfield  {author} {\bibinfo {author} {\bibfnamefont {J.~L.}\ \bibnamefont
  {Abelleira~Fernandez}}, \bibinfo {author} {\bibfnamefont {C.}~\bibnamefont
  {Adolphsen}}, \bibinfo {author} {\bibfnamefont {A.~N.}\ \bibnamefont {Akay}},
  \bibinfo {author} {\bibfnamefont {H.}~\bibnamefont {Aksakal}}, \bibinfo
  {author} {\bibfnamefont {J.~L.}\ \bibnamefont {Albacete}}, \bibinfo {author}
  {\bibfnamefont {S.}~\bibnamefont {Alekhin}}, \bibinfo {author} {\bibfnamefont
  {P.}~\bibnamefont {Allport}}, \bibinfo {author} {\bibfnamefont
  {V.}~\bibnamefont {Andreev}}, \bibinfo {author} {\bibfnamefont {R.~B.}\
  \bibnamefont {Appleby}}, \bibinfo {author} {\bibfnamefont {E.}~\bibnamefont
  {Arikan}}, \ and\ \bibinfo {author} {\bibnamefont {et~al.}},\ }\href
  {\doibase 10.1088/0954-3899/39/7/075001} {\bibfield  {journal} {\bibinfo
  {journal} {Journal of Physics G: Nuclear and Particle Physics}\ }\textbf
  {\bibinfo {volume} {39}},\ \bibinfo {pages} {075001} (\bibinfo {year}
  {2012})}\BibitemShut {NoStop}%
\bibitem [{\citenamefont {Khalek}\ \emph {et~al.}(2019)\citenamefont {Khalek},
  \citenamefont {Bailey}, \citenamefont {Gao}, \citenamefont {Harland-Lang},\
  and\ \citenamefont {Rojo}}]{LHeCPDFs}%
  \BibitemOpen
  \bibfield  {author} {\bibinfo {author} {\bibfnamefont {R.~A.}\ \bibnamefont
  {Khalek}}, \bibinfo {author} {\bibfnamefont {S.}~\bibnamefont {Bailey}},
  \bibinfo {author} {\bibfnamefont {J.}~\bibnamefont {Gao}}, \bibinfo {author}
  {\bibfnamefont {L.}~\bibnamefont {Harland-Lang}}, \ and\ \bibinfo {author}
  {\bibfnamefont {J.}~\bibnamefont {Rojo}},\ }\href {\doibase
  10.21468/scipostphys.7.4.051} {\bibfield  {journal} {\bibinfo  {journal}
  {SciPost Physics}\ }\textbf {\bibinfo {volume} {7}} (\bibinfo {year}
  {2019}),\ 10.21468/scipostphys.7.4.051}\BibitemShut {NoStop}%
\bibitem [{\citenamefont {Hou}\ \emph {et~al.}(2019)\citenamefont {Hou},
  \citenamefont {Gao}, \citenamefont {Hobbs}, \citenamefont {Xie},
  \citenamefont {Dulat}, \citenamefont {Guzzi}, \citenamefont {Huston},
  \citenamefont {Nadolsky}, \citenamefont {Pumplin}, \citenamefont {Schmidt},
  \citenamefont {Sitiwaldi}, \citenamefont {Stump},\ and\ \citenamefont
  {Yuan}}]{ct18}%
  \BibitemOpen
  \bibfield  {author} {\bibinfo {author} {\bibfnamefont {T.-J.}\ \bibnamefont
  {Hou}}, \bibinfo {author} {\bibfnamefont {J.}~\bibnamefont {Gao}}, \bibinfo
  {author} {\bibfnamefont {T.~J.}\ \bibnamefont {Hobbs}}, \bibinfo {author}
  {\bibfnamefont {K.}~\bibnamefont {Xie}}, \bibinfo {author} {\bibfnamefont
  {S.}~\bibnamefont {Dulat}}, \bibinfo {author} {\bibfnamefont
  {M.}~\bibnamefont {Guzzi}}, \bibinfo {author} {\bibfnamefont
  {J.}~\bibnamefont {Huston}}, \bibinfo {author} {\bibfnamefont
  {P.}~\bibnamefont {Nadolsky}}, \bibinfo {author} {\bibfnamefont
  {J.}~\bibnamefont {Pumplin}}, \bibinfo {author} {\bibfnamefont
  {C.}~\bibnamefont {Schmidt}}, \bibinfo {author} {\bibfnamefont
  {I.}~\bibnamefont {Sitiwaldi}}, \bibinfo {author} {\bibfnamefont
  {D.}~\bibnamefont {Stump}}, \ and\ \bibinfo {author} {\bibfnamefont {C.~P.}\
  \bibnamefont {Yuan}},\ }\href@noop {} {\enquote {\bibinfo {title} {New cteq
  global analysis of quantum chromodynamics with high-precision data from the
  lhc},}\ } (\bibinfo {year} {2019}),\ \Eprint
  {http://arxiv.org/abs/1912.10053} {arXiv:1912.10053 [hep-ph]} \BibitemShut
  {NoStop}%
\bibitem [{\citenamefont {Lees}\ \emph {et~al.}(2014)\citenamefont {Lees} \emph
  {et~al.}}]{Lees:2014xha}%
  \BibitemOpen
  \bibfield  {author} {\bibinfo {author} {\bibfnamefont {J.}~\bibnamefont
  {Lees}} \emph {et~al.} (\bibinfo {collaboration} {BaBar}),\ }\href {\doibase
  10.1103/PhysRevLett.113.201801} {\bibfield  {journal} {\bibinfo  {journal}
  {Phys. Rev. Lett.}\ }\textbf {\bibinfo {volume} {113}},\ \bibinfo {pages}
  {201801} (\bibinfo {year} {2014})},\ \Eprint {http://arxiv.org/abs/1406.2980}
  {arXiv:1406.2980 [hep-ex]} \BibitemShut {NoStop}%
\bibitem [{\citenamefont {Lees}\ \emph {et~al.}(2017)\citenamefont {Lees} \emph
  {et~al.}}]{Lees:2017lec}%
  \BibitemOpen
  \bibfield  {author} {\bibinfo {author} {\bibfnamefont {J.}~\bibnamefont
  {Lees}} \emph {et~al.} (\bibinfo {collaboration} {BaBar}),\ }\href {\doibase
  10.1103/PhysRevLett.119.131804} {\bibfield  {journal} {\bibinfo  {journal}
  {Phys. Rev. Lett.}\ }\textbf {\bibinfo {volume} {119}},\ \bibinfo {pages}
  {131804} (\bibinfo {year} {2017})},\ \Eprint
  {http://arxiv.org/abs/1702.03327} {arXiv:1702.03327 [hep-ex]} \BibitemShut
  {NoStop}%
\bibitem [{\citenamefont {Aaij}\ \emph {et~al.}(2018)\citenamefont {Aaij} \emph
  {et~al.}}]{lhcb}%
  \BibitemOpen
  \bibfield  {author} {\bibinfo {author} {\bibfnamefont {R.}~\bibnamefont
  {Aaij}} \emph {et~al.} (\bibinfo {collaboration} {LHCb}),\ }\href {\doibase
  10.1103/PhysRevLett.120.061801} {\bibfield  {journal} {\bibinfo  {journal}
  {Phys. Rev. Lett.}\ }\textbf {\bibinfo {volume} {120}},\ \bibinfo {pages}
  {061801} (\bibinfo {year} {2018})},\ \Eprint
  {http://arxiv.org/abs/1710.02867} {arXiv:1710.02867 [hep-ex]} \BibitemShut
  {NoStop}%
\bibitem [{\citenamefont {Baumgart}\ \emph {et~al.}(2009)\citenamefont
  {Baumgart}, \citenamefont {Cheung}, \citenamefont {Ruderman}, \citenamefont
  {Wang},\ and\ \citenamefont {Yavin}}]{Baumgart:2009tn}%
  \BibitemOpen
  \bibfield  {author} {\bibinfo {author} {\bibfnamefont {M.}~\bibnamefont
  {Baumgart}}, \bibinfo {author} {\bibfnamefont {C.}~\bibnamefont {Cheung}},
  \bibinfo {author} {\bibfnamefont {J.~T.}\ \bibnamefont {Ruderman}}, \bibinfo
  {author} {\bibfnamefont {L.-T.}\ \bibnamefont {Wang}}, \ and\ \bibinfo
  {author} {\bibfnamefont {I.}~\bibnamefont {Yavin}},\ }\href {\doibase
  10.1088/1126-6708/2009/04/014} {\bibfield  {journal} {\bibinfo  {journal}
  {JHEP}\ }\textbf {\bibinfo {volume} {04}},\ \bibinfo {pages} {014} (\bibinfo
  {year} {2009})},\ \Eprint {http://arxiv.org/abs/0901.0283} {arXiv:0901.0283
  [hep-ph]} \BibitemShut {NoStop}%
\bibitem [{\citenamefont {Essig}\ \emph {et~al.}(2009)\citenamefont {Essig},
  \citenamefont {Schuster},\ and\ \citenamefont {Toro}}]{Essig:2009nc}%
  \BibitemOpen
  \bibfield  {author} {\bibinfo {author} {\bibfnamefont {R.}~\bibnamefont
  {Essig}}, \bibinfo {author} {\bibfnamefont {P.}~\bibnamefont {Schuster}}, \
  and\ \bibinfo {author} {\bibfnamefont {N.}~\bibnamefont {Toro}},\ }\href
  {\doibase 10.1103/PhysRevD.80.015003} {\bibfield  {journal} {\bibinfo
  {journal} {Phys. Rev. D}\ }\textbf {\bibinfo {volume} {80}},\ \bibinfo
  {pages} {015003} (\bibinfo {year} {2009})},\ \Eprint
  {http://arxiv.org/abs/0903.3941} {arXiv:0903.3941 [hep-ph]} \BibitemShut
  {NoStop}%
\end{thebibliography}%

\end{document}